\documentstyle[12pt,epsf]{article}
\begin{document}
\begin{center}
{\Large Recoil correction to the ground state energy of hydrogenlike atoms }\\
\end{center}
\begin{center}
{V.M. Shabaev \footnote{Present address:
{\it Department of Physics, St.Petersburg State University,
 Oulianovskaya 1, Petrodvorets, St.Petersburg 198904, Russia}}}
\\
\end{center}
\begin{center}
{\it Max-Planck-Institut f\"ur Physik komplexer Systeme,
N\"othnitzer Str. 38, D-01187,
Dresden, Germany}
\end{center}
\begin{center}
A.N. Artemyev
\\
\end{center}
\begin{center}
{\it Department of Physics, St.Petersburg State University,}\\
{\it Oulianovskaya 1, Petrodvorets, St.Petersburg 198904, Russia}
\end{center}
\begin{center}
T. Beier, G. Plunien, V.A. Yerokhin \footnote{
Permanent address:
{\it Institute for High Performance Computing and Data Bases, Fontanka
118, St.Petersburg 198005, Russia}
}
, and G. Soff
\\
\end{center}
\begin{center}
{\it Institut f\"ur Theoretische Physik, Technische Universit\"at
Dresden, Mommsenstrasse 13, D-01062 Dresden, Germany}
\end{center}

PACS number(s): 12.20.-m, 31.30.Jv, 31.30.Gs
\begin{abstract}
The recoil correction to the ground state energy of hydrogenlike
atoms is calculated to all orders in $\alpha Z$ in the range
$Z$ = 1--110. The nuclear size corrections to the recoil effect
are partially taken into account.
 In the case of hydrogen,
the relativistic recoil correction beyond
the Salpeter contribution and the nonrelativistic
nuclear size correction to the recoil effect, amounts to --7.2(2) kHz.
The total recoil correction to the
ground state energy in hydrogenlike uranium ($^{238}$U$^{91+}$)
constitutes 0.46 eV.

\end{abstract}

\clearpage
\section{Introduction}

The complete $\alpha Z$-dependence formulas for the nuclear
recoil corrections to the energy levels of hydrogenlike atoms
in the case of a point nucleus
were first obtained by a quasipotential method
 [1] and subsequently rederived by different approaches
[2-4].
According to [4], the nuclear size corrections to
the recoil effect can be
partially included in these formulas
by a replacement of the pure Coulomb
potential with the potential of an extended nucleus.
The total recoil correction for a state $a$ of a hydrogenlike atom
is conveniently written as the sum of a low-order term $\Delta E_{L}$
and a higher-order term $\Delta E_{H}$ [1], where ($\hbar=c=1$)
\begin{eqnarray}
\Delta E_{L}&=&\frac{1}{2M}\langle a|[
{\bf p}^2-
({\bf D}(0)\cdot{\bf p}+{\bf p}\cdot{\bf D}(0))
]|a\rangle\,,\\
\Delta E_{H}&=&\frac{i}{2\pi M}\int_{-\infty}^{\infty}d\omega\,
\langle a|\Bigl({\bf D}(\omega)-
\frac{[{\bf p},V]}{\omega+i0}
\Bigr)\nonumber\\
&&\times
G(\omega +\varepsilon_{a})
\Bigl({\bf D}(\omega)+\frac{[{\bf p},V]}{\omega+i0}\Bigr)
|a\rangle\,.
\end{eqnarray}
Here, $|a\rangle$ is the unperturbed state of the Dirac electron
in the nuclear
potential
$V(r)$,
${\bf p}$ is the momentum operator,
 $G(\omega)=(\omega-H(1-i0))^{-1}$ is the relativistic
Coulomb Green function,
$ H=(\mbox{\boldmath $\alpha$}\cdot
{\bf p})+\beta m +V\,$,
$\alpha_{l}\;(l=1,2,3)$, $\beta$ are the Dirac matrices,
$$
D_{m}(\omega)=-4\pi\alpha Z\alpha_{l}D_{lm}(\omega)\,,
$$
and $D_{ik}(\omega,r)$ is the transverse part of the photon
propagator in the Coulomb gauge:
$$
D_{ik}(\omega,{\bf r})=-\frac{1}{4\pi}\Bigl\{\frac
{\exp{(i|\omega|r)}}{r}\delta_{ik}+\nabla_{i}\nabla_{k}
\frac{(\exp{(i|\omega|r)}
-1)}{\omega^{2}r}\Bigr\}\,.
$$
In equation (2),
the scalar product is implicit.
For point-like nuclei, $V(r)=V_{C}(r)=-\alpha Z/r$.
If extended nuclei are considered, $V(r)$ is the potential of
the extended nucleus in eq. (2) and  in
calculating $\varepsilon_{a}$, $|a\rangle$, and $G(\omega)$.
Therefore, the nuclear size corrections are
completely included in the
Coulomb part of the recoil effect. In the
one-transverse-photon part
and the two-transverse-photon part (see Ref. [4]), they
are only partially included. At least for high $Z$ we expect
that this procedure accounts for the dominant part of
the nuclear size effect since
using the extended
nucleus wave function
and the extended nucleus Green function strongly
reduces the
singularities of the integrands in (1) and (2) in the nuclear
region.

The term $\Delta E_{L}$ contains all the recoil corrections
within the $(\alpha Z)^{4}m^2/M$ approximation.
Its calculation for a point nucleus, based on the virial
relations for the Dirac equation [5-7],
yields [1]
\begin{eqnarray}
\Delta E_{L}=
\frac{m^{2}-\varepsilon_{a0}^{2}}{2M}\,,
\end{eqnarray}
where $\varepsilon_{a0}$ is the Dirac electron energy for the point
nucleus case.

$\Delta E_{H}$ contains the contribution of order
$(\alpha Z)^{5} m^2/M$ and all contributions of higher order
in $\alpha Z$. To lowest order in $\alpha Z$, this term represents the
Salpeter correction [8].
The calculation of this term
 to all orders in $\alpha Z$
was
performed  in [9,10] for the case of a point nucleus.
According to these calculations, the recoil correction
to the Lamb shift of the 1s state
in hydrogen constitutes --7.1(9) kHz,
in addition to the Salpeter term.
This value is close to the $(\alpha Z)^6m^2/M$ correction
(--7.4 kHz) found in [3] and is clearly distinct from
a recent
result for the $(\alpha Z)^6m^2/M$ correction
(--16.4 kHz) obtained in
[11].
(The  $ (\alpha Z)^6\log{(\alpha Z)}
m^2/M$ corrections
cancel each other [12,13].)
The total recoil correction to the
ground state energy in $^{238}$U$^{91+}$ was calculated in [9] to be
0.51 eV.

In this work we calculate the recoil correction to the ground state
energy of hydrogenlike atoms in the range $Z$=1--110 using the
formulas (1) and (2) employing the potential of an extended nucleus.

\section{Low-order term}
Using the virial relations for the Dirac equation in a central
field [7], the formula (1) can be transformed to
(see Appendix)
\begin{eqnarray}
\Delta E_{L}&=& \frac{m^2-\varepsilon_{a0}^2}{2M}+
\frac{1}{2M}[(\varepsilon_{a0}^2-\varepsilon_{a}^2)+
(a|(\delta V)^2|a)\nonumber\\
&&+2\alpha Z\kappa(a|\sigma_{z}\delta V/r|a)-
2\varepsilon_{a}(a|\delta V|a)\nonumber\\
&&+2(m+2\varepsilon_{a}\kappa)(a|\sigma_{z}\delta V|a)
-2\alpha Z m(a|\sigma_{x}\delta V|a)\nonumber\\
&&-4m\varepsilon_{a}(a|\sigma_{x}r\delta V|a)]\,,
\end{eqnarray}
where $\varepsilon_{a}$ and $\varepsilon_{a0}$ are the Dirac electron
 energies for an extended nucleus and the
 point nucleus, respectively,
$\kappa =(-1)^{j+l+1/2}(j+1/2)$ is the relativistic angular quantum
number,
$\delta V=V(r)-V_{C}(r)$ is the deviation of the nuclear potential from
the pure Coulomb potential,
and $\sigma_{x}$ and $\sigma_{z}$ are the Pauli
matrices. Here, the notations for the radial matrix elements from [7]
are used:
\begin{eqnarray}
(a|u|b)&=&\int_{0}^{\infty}[G_{a}(r)G_{b}(r)+F_{a}(r)F_{b}(r)]
u(r)\, dr\,,\\
(a|\sigma_{z}u|b)&=&\int_{0}^{\infty}[G_{a}(r)G_{b}(r)-F_{a}(r)F_{b}(r)]
u(r)\, dr\,,\\
(a|\sigma_{x}u|b)&=&\int_{0}^{\infty}[G_{a}(r)F_{b}(r)+F_{a}(r)G_{b}(r)]
u(r)\, dr\,.
\end{eqnarray}
$G/r=g$ and $F/r=f$ are
 the radial components of the Dirac wave function for the extended
nucleus, which are
defined by
$$
\psi_{n\kappa m}({\bf r})=
\left(\begin{array}{c}
g_{n\kappa}(r)\Omega_{\kappa m}({\bf n})\\
if_{n\kappa}(r)\Omega_{-\kappa m}({\bf n})
\end{array}\right)\;.\\
$$
The first term 
on
the right side of 
equation (4)
corresponds to the low-order recoil correction for the point
nucleus (see Eq. (3)). The second term gives the nuclear size
correction . We
calculate  this term for the 
uniformly charged 
nucleus. 
In Table I, we display the results of this calculation
for the the 1s  state. The values are expressed
in terms of the function $\Delta F_{L}(\alpha Z)$ which is
defined
by
\begin{eqnarray}
\Delta E_{L}=
\frac{m^{2}-\varepsilon_{a0}^{2}}{2M}(1+\Delta F_{L}(\alpha Z))\,.
\end{eqnarray}
In order to compare the  nuclear size correction to the low-order term
with the corresponding correction to the higher-order term (see the next 
section), in the last column of the Table I we display the value 
$\Delta P_{L}(\alpha Z)$ which is defined by
\begin{eqnarray}
\Delta E_{L}=
\frac{m^{2}-\varepsilon_{a0}^{2}}{2M}+\frac{(\alpha Z)^5}{\pi n^3}
\frac{m^2}{M}\Delta P_{L}(\alpha Z)\,.
\end{eqnarray}

Using Eq. (4), 
one easily finds for an arbitrary $n$s state
and for very low $Z$ 
($\alpha Z \ll 1$)
\begin{eqnarray}
\Delta F_{L}(\alpha Z)=\frac{1}{n}\Bigl[
-\frac{12}{5}(\alpha Z)^2 (Rm)^2
-\frac{72}{35}
(\alpha Z)^3 R m\Bigr]\,,
\end{eqnarray}
where $R=\sqrt{5/3}\langle r^2\rangle^{1/2}$ is the radius of the
uniformly charged nucleus.
The first term in (10) is a pure nonrelativistic 
 one. 
It
describes the reduced mass correction to the
nonrelativistic nuclear size effect. So, if the nuclear size
correction to the energy level is calculated using the reduced
mass, this term must be omitted in 
equation (10).
The second term, which is dominant, arises from the Coulomb part 
($ \langle a|{\bf p}^2|a \rangle/(2M)$). For the standard
parametrization of the proton form factor
\begin{eqnarray}
f(p)=\frac{\Lambda^4}{(\Lambda^2+p^2)^2}\,,
\end{eqnarray}
which corresponds to
\begin{eqnarray}
\rho (r)=\frac{\Lambda^3}{8\pi}\exp{(-\Lambda r)}
\end{eqnarray}
and
\begin{eqnarray}
V(r)=-\frac{\alpha Z}{r}\Bigl[1-\frac{1}{2}\exp{(-\Lambda r)}
(2+\Lambda r)\Bigr]\,,
\end{eqnarray}
the contribution of this term to $\Delta P_{L}$ is
\begin{eqnarray}
\Delta P_{L}'=-\frac{35}{8}\pi \frac{m}{\Lambda}\,.
\end{eqnarray}
We will see in the next section that this term cancels with
the corresponding correction to the Coulomb part of the
higher-order term. This implies
that the sum of the low-order
and higher-order contributions is more regular at $r\rightarrow 0$
than each of them  separately.
\section{Higher-order term}

To calculate the higher-order term (2) we transform it in the same
way as it was done in [9]. The final 
expressions are
given by the equations (41)-(54) of Ref. [9]
where 
the pure Coulomb potential
($V_{C}(r)=-\alpha Z/r$) in
the equations  (42) and (48) has to replaced by
the potential of the extended nucleus $V(r)$. We 
calculate these expressions
 for the uniformly charged nucleus
by using the finite basis set method with the basis functions
 constructed from B-splines [14]. The algorithm of the numerical
procedure is the same as it is described in [9].
The results of the calculation for the 1s state
are presented in the second column of the Table II. 
They are
expressed in terms of the function $P(\alpha Z)$
defined by
\begin{eqnarray}
\Delta E_{H}=\frac{(\alpha Z)^5}{\pi n^3}\frac{m^2}{M}P(\alpha Z)\,.
\end{eqnarray}
For comparison,
in the third column of this table we list the point-nucleus 
results ($P_{0}(\alpha Z)$) that are obtained by the
corresponding calculation for
$R\rightarrow 0$. These point-nucleus results are in
good agreement with our previous results from [9].
In the fourth column of the table, the difference
$\Delta P=P-P_{0}$ is listed. Finally, in the last column
the Salpeter contribution [8,15]
\begin{eqnarray}
P_{S}^{(1{\rm s})}=-\frac{2}{3}\ln{(\alpha Z)}-\frac{8}{3}\,2.984129
+\frac{14}{3}\ln{2}+\frac{62}{9}
\end{eqnarray}
is displayed.

For low $Z$ the nuclear size correction to the higher-order
 term is mainly  
due to 
the Coulomb contribution
\begin{eqnarray}
\Delta E_{H}^{(C)}&=&\frac{1}{2\pi iM}\int_{-\infty}^{\infty}d\omega\,
\frac{1}{(\omega+i0)^{2}}
(\omega)\langle a|[{\bf p},V]G(\omega+\varepsilon_{a})
[{\bf p},V]|a\rangle\,.
\end{eqnarray}
It
is comparable
with the deviation of the complete $\alpha Z$-dependence value
from the Salpeter contribution (in the case of hydrogen
$\Delta P=-0.0092(2)$  while $P_{0}-P_{S}=-0.0162(3)$).
To check this result
let us calculate the finite nuclear
size correction to the Coulomb part of the $(\alpha Z)^5 m^2/M$
 contribution.
Taken to the lowest order in 
$\alpha Z$, 
formula (17) 
yields
\begin{eqnarray}
\Delta E_{H}^{(C)}=-\frac{(2\pi)^3}{2M}|\phi_{a}(0)|^2
\int d{\bf p} \frac{\sqrt{p^2+m^2}-m}{(\sqrt{p^2+m^2}+m)^2}
\frac{p^2\tilde V^2(p)}{\sqrt{p^2+m^2}}\,,
\end{eqnarray}
where $\phi_{a}(0)$ is the non-relativistic wave function
at $r=0$
and $\tilde V(p)$ is the nuclear potential in 
the momentum representation. Using the standard parametrization of the
proton form factor
\begin{eqnarray}
\tilde V(p)=-\frac{\alpha Z}{2\pi^2 p^2}\frac{\Lambda^4}
{(\Lambda^2+p^2)^2}
\end{eqnarray}
and separating the point nucleus result from (18), 
we can write for
an $n$s state
\begin{eqnarray}
\Delta E_{H}^{(C)}=\frac{(\alpha Z)^5}{\pi n^3}\frac{m^2}{M}
(-4/3+\Delta P^{(C)})\,,
\end{eqnarray}
where
\begin{eqnarray}
\Delta P^{(C)}=-4\int_{0}^{\infty}dp\, \frac{p^2}{(\sqrt{p^2+m^2}
+m)^3}\frac{m}{\sqrt{p^2+m^2}}\Bigl[\frac{\Lambda^8}{(\Lambda^2
+p^2)^4}-1\Bigr]\,.
\end{eqnarray}
Evaluation of this integral to the lowest order in $m/\Lambda$
yields
\begin{eqnarray}
\Delta P^{(C)}=\frac{35}{8}\pi \frac{m}{\Lambda}\,.
\end{eqnarray}
As we noted above, the correction (22) cancels with the
corresponding correction to the low-order term (see Eq. (14)).
For 
$\langle r^2\rangle ^{1/2}
=0.862(12)$ fm [16], which corresponds to
$\Lambda=\sqrt{12}/\langle r^2\rangle ^{1/2}=0.845\:m_{p}=793$ MeV,
the formula (22) yields
$\Delta P^{(C)}=0.00886$
while the exact calculation of the integral (21) amounts to
$\Delta P^{(C)}=0.00874$.
These results are in good agreement with the corresponding result
($\Delta P=0.0092(2)$)
from
the Table II. 

\section{Discussion}

In this work we have calculated the  recoil correction to 
the ground state energy of hydrogenlike atoms for 
extended
nuclei 
in the range $Z=1-110$.  
This correction is conveniently written in the
form
\begin{eqnarray}
\Delta E= \frac{(\alpha Z)^2}{2M}+\frac{(\alpha Z)^5}{\pi}
\frac{m^2}{M}P_{FS}(\alpha Z)\,.
\end{eqnarray}
The function $P_{FS}(\alpha Z)=P(\alpha Z)+\Delta P_{L}(\alpha Z)$
is shown in Fig. 1. For comparison, the point nucleus function
$P_{0}(\alpha Z)$ and the Salpeter function $P_{S}(\alpha Z)$
are also presented in this figure.
The Table III displays the values of the 
recoil corrections (in eV)
in the range $Z$=10--110. 

In the case of hydrogen we find
that the  recoil correction 
amounts to 
$\Delta E= -7.2(2)$ kHz 
beyond the
Salpeter contribution and the nonrelativistic 
 nuclear size correction to the recoil effect
(the first term in Eq. (10)). It 
almost  coincides with
the point nucleus result. This is caused
by the fact that the nuclear size
correction  to the higher-order term
(Eq. (22)) and the relativistic nuclear size correction
to the low-order term (Eq. (14)) cancel each other.

For high $Z$, where the $\alpha Z$ expansion
as well as the reduced mass approximation
are not valid any more, we 
should
not separate any contributions
from the total  recoil effect.
In the case of hydrogenlike uranium ($^{238}$U$^{91+}$),
 the total recoil
correction constitutes
$\Delta E=  \Delta E_{L}+\Delta E_{H}=0.46 $ eV and
is by 10 \% smaller than the corresponding point nucleus
value ($\Delta E_{\rm p.n.} =0.51 $ eV) found in [9].
This improvement
affects the current numbers of the Lamb shift  prediction [17].

Finally, we note a very significant amount of the nuclear size
effect for $Z$=110. According to the Table III, the finite
nuclear size modifies the point nucleus result by more than 40\%.

\section*{Acknowledgements}
Valuable conversations with  S.G. Karshenboim, P. Mohr, 
K. Pachucki, and A.S. Yelkhovsky are gratefully acknowledged.
V. M. S. thanks the Institut f\"ur Theoretische Physik 
at the Technische Universit\"at Dresden
for the kind hospitality.
 The work of V. M. S., A. N. A., and V. A. Y.
was supported in part by  Grant No. 95-02-05571a from
RFBR. Also we gratefully acknowledge support by BMBF, DAAD, DFG, and
GSI. T.~B. and. G.~P. express their gratitude to the Department of
Physics at the St. Petersburg State University, where they have been
welcome in a very friendly atmosphere.

\newpage
\section*{Appendix}

Using the identity ${\bf p}^2=(\mbox{\boldmath $\alpha$}\cdot {\bf p})^2$,
the Coulomb part of the low-order term can be written as
\begin{eqnarray}
\Delta E_{L}^{(C)}&=&\langle a|\frac{p^2}{2M}|a\rangle=
\frac{1}{2M}\langle a|(\varepsilon_{a} -\beta m -V)^2|a\rangle\nonumber\\
&=&\frac{1}{2M}[\varepsilon_{a}^2+m^2+\langle a|(V^2-2\varepsilon_{a} V)|
a\rangle\nonumber\\
&&+\,2m\langle a|\beta(V-\varepsilon_{a})|a\rangle]\,.
\end{eqnarray}
As described in detail in [18], the Breit part of the
low-order term  can be transformed to
\begin{eqnarray}
\Delta E_{L}^{(B)}&=&-\frac{1}{2M}\langle a|
[{\bf D}(0)\cdot{\bf p}+{\bf p}\cdot{\bf D}(0)]|a\rangle
\nonumber\\
&=&-\frac{1}{2M}\langle a|\frac{\alpha Z}{r}
\Bigl(\mbox{\boldmath $\alpha$}+\frac{(\mbox{\boldmath $\alpha$}
\cdot {\bf r})}{r^2}\Bigr)\cdot{\bf p}|a\rangle\nonumber\\
&=&-\frac{\alpha Z}{2M}\langle a|\frac{1}{r}
\Bigl(2\varepsilon_{a}-2\beta m-2V+\frac{i \kappa}{r}
\alpha_{r}\beta\Bigr)|a\rangle\nonumber\\
&=&\frac{1}{2M}\Bigl[2\alpha Z\langle a|V/r|a\rangle
-2\alpha Z\varepsilon_{a} \langle a|1/r|a\rangle \nonumber\\
&&+\,2\alpha Z\langle a|m\beta/r|a\rangle+
2\kappa \alpha Z\int_{0}^{\infty}g_{a}f_{a}dr\,,
\end{eqnarray}
where $\alpha_{r}=(\mbox{\boldmath $\alpha$}\cdot {\bf r})/r$ and
$\kappa =(-1)^{j+l+1/2}(j+1/2)$ is the relativistic angular quantum
number of the state $a$.  In 
the following
we will use the notations of Ref. 
[7],
\begin{eqnarray}
A^s&=&\int_{0}^{\infty}(G^2+F^2)r^{s}dr\,,\\
B^s&=&\int_{0}^{\infty}(G^2-F^2)r^{s}dr\,,\\
C^s&=&2\int_{0}^{\infty}GFr^{s}dr\,,
\end{eqnarray}
where $G/r=g$ and $F/r=f$ are the radial components of the
Dirac wave function for the extended nucleus,
and the radial scalar product defined by the equations (5)-(7).
Using the equation (2.9) of Ref. [7], we find
\begin{eqnarray}
\Delta E_{L}&=&\frac{1}{2M}[\varepsilon_{a}^2-m^2+(a|\delta V(\delta V-
2\varepsilon_{a})|a)\nonumber\\
&&-\alpha Z(\alpha Z A^{-2}-\kappa C^{-2}-2m B^{-1})]\,,
\end{eqnarray}
where $\delta V=V-V_{C}=V+\alpha Z/r$.
>From the equations (2.8)-(2.10) of Ref. [7], one obtains
\begin{eqnarray}
\alpha Z A^{-2}-\kappa C^{-2}&=&-2\kappa (a|\sigma_{z}\delta V/r|a)
+2m(a|\sigma_{x}\delta V|a)\,,\\
2\alpha Z m B^{-1}&=&2(m^2-\varepsilon_{a}^{2})+2(m+2\varepsilon_{a}\kappa)
(a|\sigma_{z}\delta V|a)\nonumber\\
&&-4m\varepsilon_{a} (a|\sigma_{x}r\delta V|a)\,.
\end{eqnarray}
Substituting (30) and (31) into (29), we find
\begin{eqnarray}
\Delta E_{L}&=&\frac{m^2-\varepsilon_{a}^2}{2M}+
\frac{1}{2M}\Bigl[(a|(\delta V)^2|a)+2\alpha Z \kappa
(a|\sigma_{z}\delta V/r|a)\nonumber\\
&&-2\varepsilon_{a} (a|\delta V|a)+2(m+2\varepsilon_{a} \kappa)
(a|\sigma_{z} \delta V|a)\nonumber\\
&&-2\alpha Z m (a|\sigma_{x}\delta V|a)-4m\varepsilon_{a}
(a|\sigma_{x}r\delta V|a)\Bigr]\,.
\end{eqnarray}
Separating the point nucleus result 
from the right side of (32),
we get the equation (4).

\newpage
\begin{table}
\caption{
Nuclear size correction to the
low-order term for the 1s state 
expressed in terms of the functions $\Delta F_{L}(\alpha Z)$
and $\Delta P_{L}(\alpha Z)$,
 defined by equations (8) and (9), respectively.
The values of the nuclear radii employed in the calculation
are taken from [16,19-23].}
\vspace{0.5 cm}
\begin{tabular}{|c|l|l|l|}  \hline
$Z$&$\langle r^2\rangle ^{1/2}$, fm
&$\Delta F_{L}(\alpha Z)$&$\Delta P_{L}(\alpha Z)$\\ \hline
1&0.862&-0.337$\times 10^{-8}$&-0.0136\\ \hline
2&1.673&-0.519$\times 10^{-7}$&-0.0262\\ \hline
5&2.397&-0.102$\times 10^{-5}$&-0.0329\\ \hline
10&3.024&-0.976$\times 10^{-5}$&-0.0394\\ \hline
20&3.476&-0.933$\times 10^{-4}$&-0.0472\\ \hline
30&3.928&-0.406$\times 10^{-3}$&-0.0607\\ \hline
40&4.270 &-0.126$\times 10^{-2}$&-0.0797\\ \hline
50&4.655&-0.340$\times 10^{-2}$&-0.1099\\ \hline
60&4.914&-0.823$\times 10^{-2}$&-0.1539\\ \hline
70&5.317&-0.0195&-0.2295\\ \hline
80&5.467&-0.0436&-0.3442\\ \hline
90&5.802&-0.0993&-0.5506\\ \hline
92&5.860&-0.117&-0.6073\\ \hline
100&5.886&-0.224&-0.9038\\ \hline
110&5.961&-0.517&-1.572\\ \hline
\end{tabular}
\end{table}

\begin{table}
\caption{
Higher-order term  for the 1s state 
expressed in terms of the function $P(\alpha Z)$ defined by Eq. (15).
The nuclear radii employed in the calculation
 are the same as in Table I.
$P_{0}(\alpha Z)$ is the related value for the point nucleus
and $\Delta P=P-P_{0}$.
 $P_{S}(\alpha Z)$ is the
Salpeter contribution obtained by Eq. (16).}
\vspace{0.5 cm}
\begin{tabular}{|c|l|l|l|l|}  \hline
$Z$&$P(\alpha Z)$&$P_{0}(\alpha Z)$&$\Delta P(\alpha Z)$&
$P_{S}(\alpha Z)$\\ \hline
1&5.4391(3)&5.4299(3)&0.0092(2)&5.4461\\ \hline
2&4.9703(3)&4.9528(3)&0.0175(2)&4.9840\\ \hline
5&4.3281(3)&4.3034(3)&0.0247(2)&4.3731\\ \hline
10&3.828&3.795&0.031&3.9110\\ \hline
20&3.330&3.294&0.036&3.4489\\ \hline
30&3.086&3.044&0.043&3.1786\\ \hline
40&2.977 &2.927&0.050&2.9868\\ \hline
50&2.973&2.914&0.060&2.8380\\ \hline
60&3.072&3.006&0.066&2.7165\\ \hline
70&3.295&3.234&0.061&2.6137\\ \hline
80&3.686&3.672&0.013&2.5247\\ \hline
90&4.330&4.521&-0.191&2.4462\\ \hline
92&4.501&4.779&-0.277&2.4315\\ \hline
100&5.40&6.41&-1.01&2.3759\\ \hline
110&7.24&12.43&-5.19&2.3124\\ \hline
\end{tabular}
\end{table}
\begin{table}
\caption{
Recoil corrections in eV. For comparison,  
the nonrelativistic recoil correction is given separately. The last
column displays the deviation from the point nucleus results
for the total recoil effect.
The mass values are given in nuclear mass units. They
were taken from [24], except for
$Z=110$ where we adopted the value of [21]. }
\vspace{0.5 cm}
\begin{tabular}{|r|r|l|l|l|}  \hline
 $Z$ & $M/A$ & nonrel. recoil & total recoil & finite size effect \\ \hline
  10 & 20.2 & 0.037 & 0.037 & \\ \hline
  20 & 40.1 & 0.075 & 0.075 & \\ \hline
  30 & 65.4 & 0.104 & 0.105 & \\ \hline
  40 & 91.2 & 0.134 & 0.137 & \\ \hline
  50 & 118.7 & 0.163 & 0.171 & \\ \hline
  60 & 144.2 & 0.196 & 0.215 & --0.001 \\ \hline
  70 & 173.0 & 0.227 & 0.269 & --0.003 \\ \hline
  79 & 197.0 & 0.26 & 0.33 & --0.01 \\ \hline
  80 & 200.6 & 0.26 & 0.34 & --0.01 \\ \hline
  82 & 207.2 & 0.27 & 0.36 & --0.01 \\ \hline
  90 & 232.0 & 0.30 & 0.44 & --0.03 \\ \hline
  92 & 238.0 & 0.30 & 0.46 & --0.05 \\ \hline
 100 & 257.1 & 0.34 & 0.61 & --0.14 \\ \hline
 110 & 268.0 & 0.42 & 0.97 & --0.75 \\ \hline
\end{tabular}
\end{table}

\clearpage

\begin{figure}
\centerline{\mbox{\epsfxsize=13.6cm\epsffile{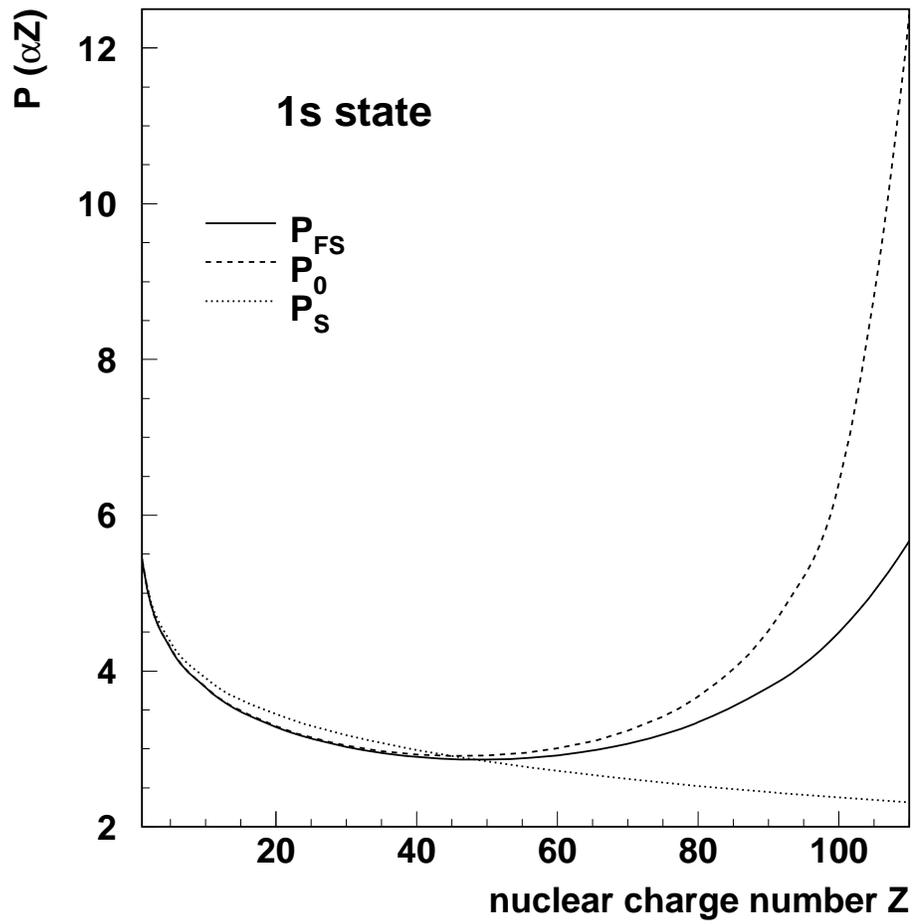}}}
\caption{The function $P_{FS}(\alpha Z)$, compared to
$P_{0}(\alpha Z)$ and $P_{S}(\alpha Z)$. }
\end{figure}

\end{document}